**Perspective: Challenges and opportunities for high-quality battery production at scale**

Peter M. Attia*, Eric Moch, Patrick K. Herring

Glimpse, Somerville, MA 02143

**Abstract**

As the impacts of climate change become increasingly apparent, the need for widespread electrification is now internationally recognized. As a result, global battery production is set to surge over the next decade. Unfortunately, however, batteries are both difficult to produce at the gigawatt-hour scale and sensitive to minor manufacturing variation. As a result, the battery industry has already experienced a number of both highly-visible safety incidents and under-the-radar reliability issues—a trend that will only worsen if left unaddressed. In this perspective, we highlight both the challenges and opportunities to enable battery quality at scale. We first describe the interplay between various battery failure modes and their numerous root causes. We then discuss the tensions at play to manage and improve battery quality during cell production. We hope our perspective brings greater visibility to the battery quality challenge to enable safe global electrification.

**\*Corresponding author**: peter@glimp.se

We currently live in exciting times for the battery industry. In light of the increasingly visible impacts of climate change[1], consumer, corporate, and governmental support for electric vehicles (EVs) and stationary energy storage is crescendoing.[2,3] The industry is projected to grow by 30% per year until 2030.[4] A planetary-scale energy transition is well underway, requiring unprecedented volumes of battery-powered energy storage.

However, the global battery production ramp is threatened by looming challenges. While concerns around the materials supply chain and the "valley of death" for new battery concepts are now well-established[5,6], we highlight an equally important but less frequently discussed risk: poor battery quality and its impacts on battery safety, reliability, and the financial success of cell manufacturing (Figure 1). Indeed, since the commercialization of lithium-ion battery technology in 1991,[7,8] several high-profile safety events (Figure 1a) have occurred in sectors such as consumer electronics, electric micromobility, EVs, aviation, and medical devices.[9,10] One infamous EV safety case required a $1.9B fleetwide recall.[11,12] Unfortunately, as applications of battery technology have proliferated, the incidence rate of catastrophic events appears to be increasing; for instance, hundreds of electric bike fires have occurred over the past few years in New York City alone, triggering legislation[13]. Underestimating these battery safety risks will have devastating financial, human, and environmental costs and jeopardize consumer and regulatory confidence in battery technology.

**a.** Safety

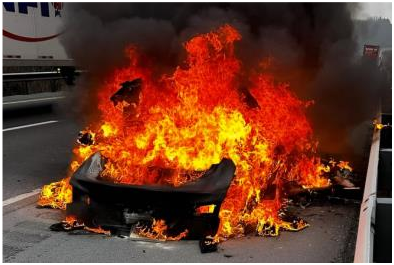

**b.** Reliability

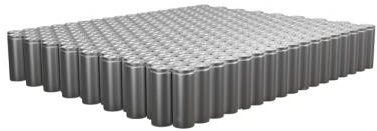

**c.** Manufacturability

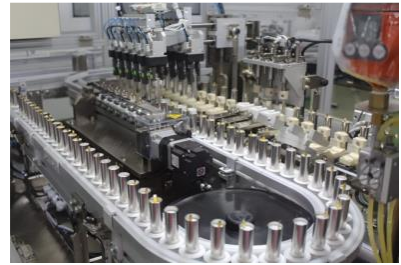

**Figure 1**. Three challenges related to poor battery quality. (a) Safety events, in which a single battery defect can cause harm to humans or the environment. (b) Pack/device reliability, in which a single cell can cause an entire pack or device to fail. Failure within warranty will require replacement at the expense of the manufacturer. (c) Manufacturability, or the immense difficulty of producing batteries at scale profitably (i.e., with high yield, throughput, and ramp up time) and while maintaining high quality (i.e., while maintaining high purities and tight tolerances).

Aside from headline-grabbing safety events, battery quality issues can have outsize impacts on the reliability of battery-powered devices (Figure 1b). For instance, an EV pack consists of many cells arranged in series and in parallel, often combined into modules. (Although we focus on EVs in this article, these principles apply to any battery-powered device). These configurations determine the sensitivity of the pack to cells that exhibit either open-circuit failure, short-circuit failure, or even heterogeneity. A cell that exhibits open-circuit failure will cause all cells connected to it in series to become inoperable (the "Christmas tree light" effect).[14–16] A cell that exhibits short-circuit failure will cause all cells connected to it in series to overcharge[17] and cause all cells connected to it in parallel to discharge into it in order to maintain the same voltage[14,17]—not to mention the risk of short-induced thermal runaway propagating to neighboring cells[18–20]. Additionally, but less critically, variability in initial cell energy and cell aging can cause a pack to underperform relative to an otherwise identical pack with no cell-to-cell variability.[21–23] While these issues can be partially managed by hardware- and software-based pack balancing strategies, these approaches add mass and complexity to pack design and are often insufficient to manage severe issues.[15,16] In summary, given the punishing physics of battery pack reliability, the failure of a single cell can cause complete pack failure.

The core challenge underlying these safety and reliability issues is the unforgiving requirements of battery production at scale (Figure 1c): namely, high production yields and throughputs along with extreme tolerance and purity specifications. A 38 GWh/year battery gigafactory produces a staggering six million cells per day—or nearly 70 cells per second.[24] Simultaneously, modern batteries must be manufactured to geometric tolerances on the order of a few microns while avoiding a host of similarly-sized particle contaminants.[25–27] Ensuring that each cell meets these specifications while being manufactured at high rates across multiple production lines that are continuously modified, maintained, and upgraded is a colossal task.[6,24,28] Thus, quality control is a key differentiator between a top-tier and bottom-tier cell producer. Ultimately, given the inevitable quality obstacles faced during large-scale battery production, both cell producers and original equipment manufacturers (OEMs; in this context, the manufacturers of any battery-powered device) must learn to understand and manage this issue.

Based on our experiences in the battery industry, we believe ensuring battery quality at scale is perhaps the most important technical challenge hindering the ability to rapidly ramp battery production in the years to come. Yet a comprehensive framework and vision for quality in the battery industry is in its infancy. In this perspective, we share our outlook on the challenges and opportunities for enabling battery quality at scale. First, we define battery quality and its relationship to other key attributes such as lifetime, failure, safety, reliability, and manufacturing performance. We then outline the difficult options available for managing and improving battery quality during cell production. Finally, we chart a course for survival in this formidable environment. We hope our perspective sheds light on this critical challenge and catalyzes a broader acceptance of its importance for an electrified future.

## Defining key attributes related to battery quality

Fundamentally, the challenge of ensuring battery quality is driven by the complexity of battery performance. An especially important, sensitive, and complex pillar of battery performance is battery lifetime and failure. While our collective understanding of this topic has continued to grow, we also continue to learn how complex and interdependent battery lifetime and failure can be.[29–31] Note that although we focus on lithium-ion batteries throughout this perspective, these principles apply to all electrochemical batteries (e.g., lithium-ion, lithium-metal, sodium-ion, and more; aqueous and non-aqueous; primary and secondary).

In Figure 2, we present a taxonomy of battery failure. We first list influential factors that drive battery failure (Figure 2a). We define three categories of battery failure, ordered by severity: *performance degradation* (Figure 2b), *functional failure* (Figure 2c), and *safety events* (Figure 2d). We discuss key terms and concepts embodied in this figure in the remainder of this section.

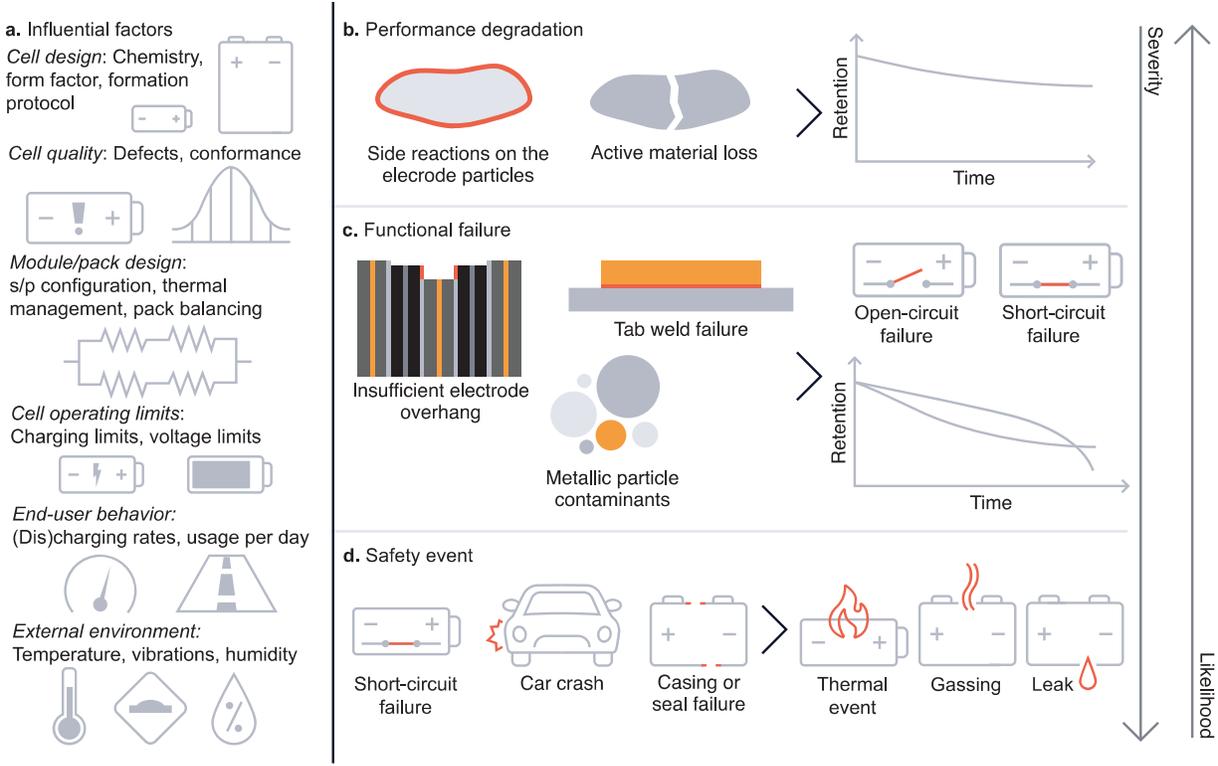

**Figure 2.** A taxonomy of battery failure. (a) Factors that influence battery lifetime and failure. The cell producer or OEM can control some of these factors but not all. (b)–(d) Three categories of battery failure, in order of severity and increasing order of likelihood (from top to bottom). These three categories of failure influence and interact with each other. (b) Performance degradation. Performance degradation is often caused by (electro)chemical side reactions that consume lithium or inactivate electrode material, resulting in diminished cell performance. (c) Functional failure. Functional failures are often caused by electrochemical, mechanical, or contamination issues (e.g., lithium plating, weld failure, and metallic contaminants) that cause cells to exhibit open-circuit failure, short-circuit failure, or severely diminished performance. (d) Safety events. Safety events can be initiated by a variety of root causes, most notably internal shorting but also external initiation events (e.g., a car crash) and casing/seal failure[32], and result in an event that harms the health of humans and/or the environment.

***Classifying battery failure and defining key terms.***—We begin by describing three broad classifications of battery failure and define key concepts along the way. While we focus on lithium-ion batteries throughout this work, our concepts apply equally to other battery chemistries such as lithium-metal and sodium-ion as well.

*Performance degradation* (Figure 2b).—As any user of a battery-powered device has likely experienced, the available energy of a battery typically decreases with time.[29–31,33,34] Other aspects of performance, such as rate capability, often decrease as well.[30,31,35] Generally, the root causes of this *performance degradation* are electrochemical and chemical degradation modes (subsequently referred to as "electrochemical" for simplicity) and have been the focus of much of the literature on battery lifetime.[29–31] Two classic electrochemical degradation modes include solid-electrolyte interphase growth[33,36–38], which consumes lithium inventory and decreases battery capacity and energy, and cathode-electrolyte interface growth[39], which increases cell internal resistance. Another significant performance degradation mode is active material loss from the positive and negative electrodes, in which electrode host sites become inaccessible for lithium ions.[40,41] However, many subtle effects, including current collector corrosion[42,43] and "cross talk" between the electrodes[44,45], can also contribute.[30,31]

Nearly all batteries experience performance degradation to some degree, and minimizing its extent is critical to improve battery sustainability and to bring next-generation battery chemistries to market.[46] Furthermore, the long duration of electrochemical lifetime testing is a major bottleneck to innovation in battery technology.[47,48] However, generally speaking, some performance degradation over life is considered acceptable by end users. Additionally, energy retention can be optimized to a remarkable extent today, as demonstrated by the Dahn lab's work on "million-mile batteries"[48] and beyond[49]. Overall, while performance degradation is certainly a key element of lifetime and failure, this category of battery failure does not threaten global electrification efforts to the same extent as the two categories that follow.

*Functional failure* (Figure 2c).—A second category of battery failure, subsequently termed *functional failure*, is a broader class of failure that renders the cell unable to meet its functional requirements. In other words, the cell becomes inoperable or, at minimum, exhibits severely diminished utility. Broadly, *open-circuit* and *short-circuit* failures are two major classes of functional failure. In a cell context, open-circuit failure refers to a cell with a broken electronic pathway (very high resistance), and short-circuit failure refers to a cell with an inadvertent electronic connection between the electrodes (very low resistance). These failure modes are internal to the cell, such as an internal short. Unlike most cases of performance degradation, these two issues can render the cell and/or pack entirely inoperable. For instance, an internal short can inhibit charging and impact pack balancing.[14,16,17] In these cases, failure is often more clear but still arbitrary (e.g., what short-circuit current should be considered failing?). We discuss the root causes of these failure modes in detail at a later point.

Performance degradation is often codified as functional failure by specifying limits for its extent. For instance, some commonly cited limits for energy retention in publications and EV warranties are 70%[50] or 80%[51]. Of course, these limits are arbitrary, as a battery is only marginally less useful with 79.9% energy retention than with 80.1% energy retention. Conventionally, the terms *battery lifetime*, *cycle life*, and *calendar life* implicitly refer to functional requirements related to performance degradation. For example, a calendar life requirement might be that a cell retains 80% of its energy over a period of eight years. Severe electrochemical events, such as a "knee" in capacity or energy retention[34] or an "elbow" in internal resistance[35], are major threats to meeting these functional requirements. Avoiding knees and elbows may also be considered functional requirements in their own right, as they can lead to user frustration relative to "graceful failure" scenarios[52] and are often correlated with other functional failures (e.g., lithium plating may be a root cause of both knees and internal shorting[34,53]).

A key term in discussing functional failures is *reliability*, which is defined as how well a product can perform its intended functions given a specified set of operating conditions and a specified period of time.[54] Thus, *battery reliability* can be defined as how well a battery avoids functional failure over its desired operating lifetime given a set of operating conditions. As previously discussed, single-cell reliability is a key determinant of pack reliability. We illustrate how single-cell reliability translates to pack-level reliability in Figure 3. Reliability science is well-established[55] and can help practitioners empirically test for functional failure modes[56,57], but the large number of operating conditions, wide variety of failure modes, and high lifetime requirements make battery reliability testing challenging. While accelerated testing (e.g., high temperature or fast cycling)[58,51] and early prediction techniques (e.g., physics- or data-driven techniques)[47,59] can help reduce the cost of reliability testing, the discrete, latent nature of many functional failures (e.g., a sudden tab weld failure[60]) sets a lower bound on the cost reduction. In other words, testing for "threshold" failure mechanisms[34] for which the threshold is unknown (i.e., when the tab weld will fail) is limited by the time it takes to identify the threshold, by definition. Similarly, electrochemomechanical interactions between performance degradation and functional failures (e.g., gassing due to side reactions and/or electrode swelling[34,61,62]) can necessitate long testing times.

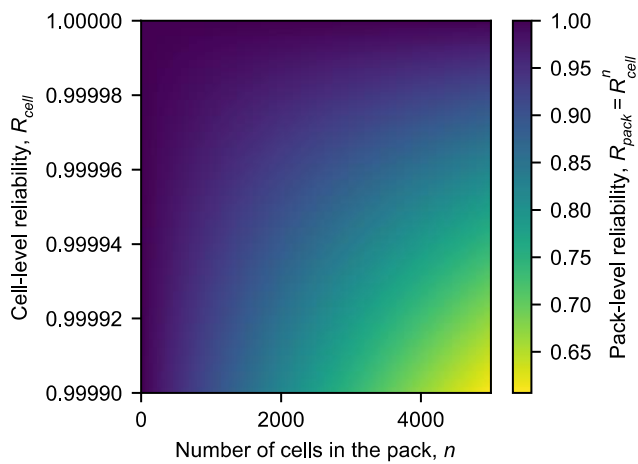

**Figure 3.** The relationship between cell-level reliability and pack-level reliability.[14,20] Here, we assume that the failure of a single cell causes the failure of the entire pack; the validity of this assumption depends on the type of failure, the pack architecture, and other factors. Under this assumption, the pack-level reliability is given as $R_{pack} = R_{cell}^n$, where $R_{pack}$ is the pack-level reliability, $R_{cell}$ is the cell-level reliability, and $n$ is the number of cells in the pack. While this visualization may suggest that packs with smaller numbers of cells are more reliable, we note that $R_{cell}$ may be lower for higher-energy cells (which are needed to build an equivalently-sized pack).

*Safety events* (Figure 2d).—The most severe category of battery failure are safety events. Here, we define a safety event as any battery issue that could cause harm to humans or the environment. With this definition, fires and explosions induced by thermal runaway are perhaps the most vivid and catastrophic issues, but the release of toxic gas via a gassing event[10] or the release of toxic liquids via a leak[18,32] would also qualify. Of course, battery safety can cause life-threatening injury and is currently a hot-button issue among legislators and regulators.[13]

As many excellent reviews of battery safety have been published[10,18,20,63], we provide a cursory review here. In general, thermal runaway can have extrinsic or intrinsic triggering mechanisms.

Extrinsic triggering mechanisms include electrical (e.g., overcharge), thermal (e.g., sudden rise in environmental temperature), and mechanical (e.g., a vehicle crash) mechanisms. Zhang et al.[64] and Lai et al.[65] have comprehensively reviewed the most significant intrinsic triggering mechanism, internal short circuiting, which can lead to excess internal heat generation and thus initiate a combustion event. However, internal shorting is not the only triggering mechanism for runaway; localized current increases (e.g., from tab tears) can create thermal "hotspots"[60], and Liu et al.[66] reported a thermal runaway mechanism driven by crosstalk. Note that aged cells with performance degradation in the absence of lithium plating generally exhibit less severe thermal runaway.[61,67] Thus, safety events are often linked to performance degradation and (especially) functional failures. The time between the initiation of a functional failure such as an internal short and the initiation of a safety event such as runaway can vary significantly.[53,64,65]

A major factor in battery design is the extent of safety margin for both intrinsic and extrinsic triggering mechanisms. For instance, separators are often designed with safety considerations such as shutdown temperature in mind[68], and the wall thickness of the casing can be tuned for resistance to side rupture and external mechanical initiation events.[69] Safety concerns also influence module and pack design.[18,19] This safety factor must consider the application; for instance, pacemaker batteries should be designed with a large safety margin on principle, while a grid storage installation intended to operate far from human habitation may require a less strict safety margin. These design choices often require tradeoffs between cell performance (e.g., energy density, rate capability, and cost) and safety margin.

*Battery quality.*—The large variety of driving forces for battery failure (Figure 2a)—and their interactions—adds another dimension to this challenge. We review these factors in Supplementary Discussion 1. Overall, the variety of factors which influence battery lifetime and failure, coupled with the wide range of failure modes, leads to hundreds or even thousands of risk factor-failure mode combinations to consider. Cell producers and OEMs can control some of these risk factors (e.g., cell design and cell operating limits) but not all (e.g., end-user behavior and cell operating environment). A deep understanding of both end-user behavior (distributions and the most extreme behaviors) and the environmental conditions introduced by end users can enable OEMs to better estimate warranty liabilities caused by these external risk factors. Here, field telemetry can play a crucial role.[70]

One of the most influential factors for battery lifetime and failure is battery quality, which underlies our entire discussion thus far. We define battery quality via one of two definitions: (a) defect rate and (b) conformance.

*Defect rate.*—Generally speaking, a poor-quality product has an unacceptably high rate of manufacturing defects. We previously discussed how battery defects cause functional failures (typically open-circuit or short-circuit failure) or safety events immediately or in use. While some of these defects are obvious, many are subtle and only manifest under special conditions. For instance, the Chevrolet Bolt safety issues were attributed to the simultaneous presence of a torn negative electrode and a folded separator.[11] Cells with poor material quality that are otherwise well-built can also be considered defective (e.g., corrosion).[71] In reality, a cell's survival is threatened by dozens of failure modes.

As shown in Figure 4, a large ensemble of battery defects can cause either open-circuit (Figure 4a) or short-circuit (Figure 4b) failure. In general, open-circuit failure is most likely to occur at the most prominent components of the electronic pathway of the cell. For instance, a failure of the weld between the tab and the terminal, a tab tear, and corrosion can all lead to open-circuit failure.[60] Furthermore, some cell safety devices, such as current interrupt devices, can intentionally cause open-circuit failure upon activation.[72] In contrast, because short-circuit failure can occur at any location where the positive and negative electrode are in electronic contact, any number of subtle, micron-scale imperfections can lead to short-circuit failure. A classic example is lithium plating, in which lithium metal "dendrites" puncture the separator and create an electronic connection between the electrodes.[53,73] While plating can occur for a number of reasons, a local region where the ratio of negative electrode to positive electrode capacity ("$n{:}p$ ratio") is less than one is often responsible.[34,53,73] This condition can occur due to electrode "overhang" issues during winding or stacking, as well as electrode coating defects.[53,74–76] Beyond plating, any number of mechanical imperfections can induce an internal short. Metallic particle contaminants, often only tens of microns in diameter, can either cause a direct short between the electrodes via separator puncture or induce metal deposition on an electrode, which can subsequently develop into a short.[26,27,76] Finally, many other defects such as separator pinholes, separator misalignment, folded separators, electrode wrinkles, jellyroll buckling, metallic burrs or tears on current collectors or tabs, and overlapping tabs can induce an internal short circuit.[11,60,62,64,65,76] Thus, internal short circuits are of particular concern in the battery industry due to the abundance of micron-scale root causes and their severity.[64,65,76]

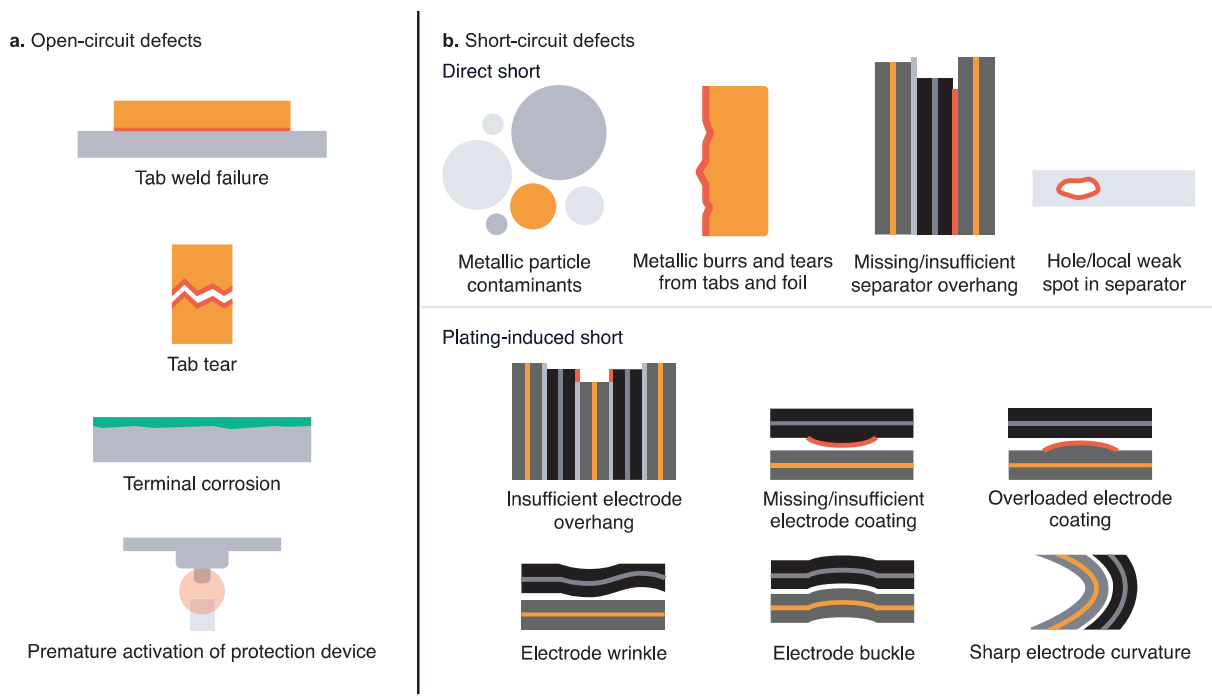

**a.** Open-circuit defects

Tab weld failure

Tab tear

Terminal corrosion

Premature activation of protection device

**b.** Short-circuit defects

Direct short

Metallic particle contaminants

Metallic burrs and tears from tabs and foil

Missing/insufficient separator overhang

Hole/local weak spot in separator

Plating-induced short

Insufficient electrode overhang

Missing/insufficient electrode coating

Overloaded electrode coating

Electrode wrinkle

Electrode buckle

Sharp electrode curvature

**Figure 4.** Common battery defects that can cause functional failures, specifically (a) open-circuit and (b) short-circuit failure. These defects can occur during manufacturing or, in the case of latent defects, over life. The relevant length scale for most of these defects is 10–100 μm. The defects depicted here are far from exhaustive.

Critically, many of these defects are latent defects, meaning they are initially present but dormant and may activate at some point over life. By definition, latent defects have no electrochemical signature until the defect manifests into failure (e.g., an internal short); in fact, even an internal short may not be detectable until its magnitude develops beyond some critical value (i.e., the point at which this localized electrochemical signal is significant enough to be detected in the "background" of the rest of the cell). We illustrate this concept in Supplementary Figure 1. Not unlike a malignant tumor, a small defect may cause premature failure for an otherwise entirely healthy cell (i.e., a cell with minimal performance degradation). This process can be thought of as a "threshold" mechanism[34], where failure occurs once some aspect of the cell's internal state has crossed some threshold. These threshold failure mechanisms can be driven by a multitude of forces (e.g., electrochemomechanical electrode swelling, gassing due to side reactions, and/or stress on a critical component) which are generally related to performance degradation mechanisms.[61,62,77,78] Both the rate of change for internal state and the magnitude of the threshold will vary based on the previously discussed factors, namely cell design, cell quality, module/pack design, cell operating limits, end-user behavior, and environmental conditions.

The key metric for this definition of battery quality is defective parts per million (DPPM).[79] However, many subtleties can make DPPM quantification difficult, if not impossible. First, developing a comprehensive understanding and precise definition of all possible defects is challenging, especially given the aforementioned interactions between factors for battery failure as well as the interactions between defects themselves[11]. For example, defective cells that may fail for one end user may not fail for another end user based on their behavior. Second, many of these defects are latent, which means they can be difficult to detect in production. Third, some fraction of these defects will escape the production line simply due to imperfect detection techniques and sensitivities.

Counterfeit cells are perhaps an extreme case of defectiveness, in which standard safety features are removed[71]. Counterfeit cells of course often have very poor quality, and many of the highly-publicized battery safety events are a result of low-quality and/or counterfeit batteries.[10,13,71]

*Conformance.*—Conformance refers to how well a manufactured product conforms to its design.[25] The battery industry often refers to nonconformance as "cell-to-cell variability".[22,56,57,80] Conformance is a broader definition of cell quality than defect rate in that defects are one instance of divergence from the intended design. In general, conformance is a function of process and quality control. Beck et al.[80] reviewed the primary drivers of nonconformance in batteries and battery production.

Lack of conformance to the design may not directly cause failure; for instance, a key quality indicator such as the distribution of cell energy may be larger than desired but still fall within an acceptable band. That said, poor conformance can influence failure in multiple ways:

- In general, poor conformance indicates poor process control, which indicates that a production line is at higher risk of producing defective cells.

- All three categories of battery failure are often highly sensitive to small differences in cell structure and composition, so small deviations may result in a significant increase in the likelihood and severity of failure and thus higher warranty exposure.[22,34,56,81,82] For instance, some proposed knee pathways exhibit superlinear sensitivities to small variations in percolation network connectivity or electrolyte additive concentration.[34] The aforementioned Chevrolet Bolt safety issue only occurred when two rare defects occurred in the same cell.[11]

- On a related note, cell testing for performance degradation, functional failures, and safety is more expensive with higher cell-to-cell variability.[81–83]

- A high degree of cell variability within a module or pack has a number of intertwined impacts on pack behavior. This issue may prevent the pack from meeting its requirements (e.g., energy or rate capability) since packs are generally limited by their weakest cell.[16,23,57] In other words, all else being equal, a pack with high variability in cell energy will have lower effective energy than a pack with low variability in cell energy. Cell variability can also cause voltage or current imbalance, which further limits performance and can cause pack-level failure as previously discussed.[14,16,17,57] Finally, these imbalances can cause heterogeneous aging under certain conditions, i.e., a cell with low energy may experience a higher effective C rate relative to nominal.[57,84,85]

- Root causing test and field failures is more difficult, as cell quality adds yet another set of complex factors to untangle.

In summary, both senses of battery quality are critical determinants of battery failure and thus the financial success of cell and EV production endeavors. We revisit battery quality in the second section of this paper.

***Manufacturing performance.***—Ultimately, of course, a business cannot build and operate a multi-billion-dollar battery factory without a return on its investment. This return is determined by a number of cell production indicators, such as yield, ramp up time, utilization, throughput, profitability, and the rate of issues in the field.[6,28] Here, we use the term "manufacturing performance" to broadly describe these performance metrics. Given the thin profit margins (often 2-3%)[86] with which battery factories operate, quality concerns are often in tension with these manufacturing performance indicators. For instance, the decision of what to do with a batch of cells with marginal failures might be heavily debated between production and quality teams. Furthermore, an engineering team may require a couple of weeks to assess the risk of a potential quality issue, but a production team must often make decisions on daily or even hourly timescales to avoid inventory buildup or, worse, a line shutdown. However, allowing defective cells to escape the factory carries significant reputational risk for both the cell producer and OEM and may require substantial engineering resources to resolve in the future.

One underappreciated attribute of manufacturing performance is dynamicism, or ability to respond to change. In an overly idealized view, a battery factory statically maintains fixed operational objectives. In reality, a factory must dynamically respond to a variety of internal (e.g., new equipment, new process learnings, new cell designs, business objectives, etc.) and external (e.g., improved or less expensive materials and components, new learnings from the field, market demand, policy incentives, etc.) factors. While too many simultaneous demands can threaten production stability, dynamicism is a key ingredient of manufacturing success.

Finally, we mention that the sustainability of battery production is becoming an increasingly important manufacturing performance metric. For instance, an estimated 30–65 kWh are consumed in the factory for every kWh of cells produced.[46,87] Furthermore, scrap rates can range from <5% to as high as 90% during ramp-up;[28,88,89] while recycling these scrap materials can improve the sustainability of battery production, better yet is to reduce the rate of scrap in the first place. Generally speaking, a strong emphasis on quality and quality control can be a powerful lever to minimize wasted material and energy during battery production.

## Managing battery quality in production

Given the frequency, severity, and inevitability of battery quality issues, both battery producers and manufacturers of battery-containing products must *manage* battery quality. Quality control often involves difficult choices made under high uncertainty, but these decisions must be made to avoid the potentially devastating risks of inaction.

In Figure 5, we propose four pathways for managing battery quality in production. These approaches are derived from process capability analysis, which is commonly employed in manufacturing environments.[90] While each of these pathways is individually presented and discussed, an "all-of-the-above" approach is often required in practice. Throughout this section, we use the example of electrode overhangs (subsequently referred to as simply "overhang") as a canonical example of a battery quality issue. Insufficient overhang may cause lithium plating, which may cause an internal short and, in extreme cases, thermal runaway.[53,74,75]

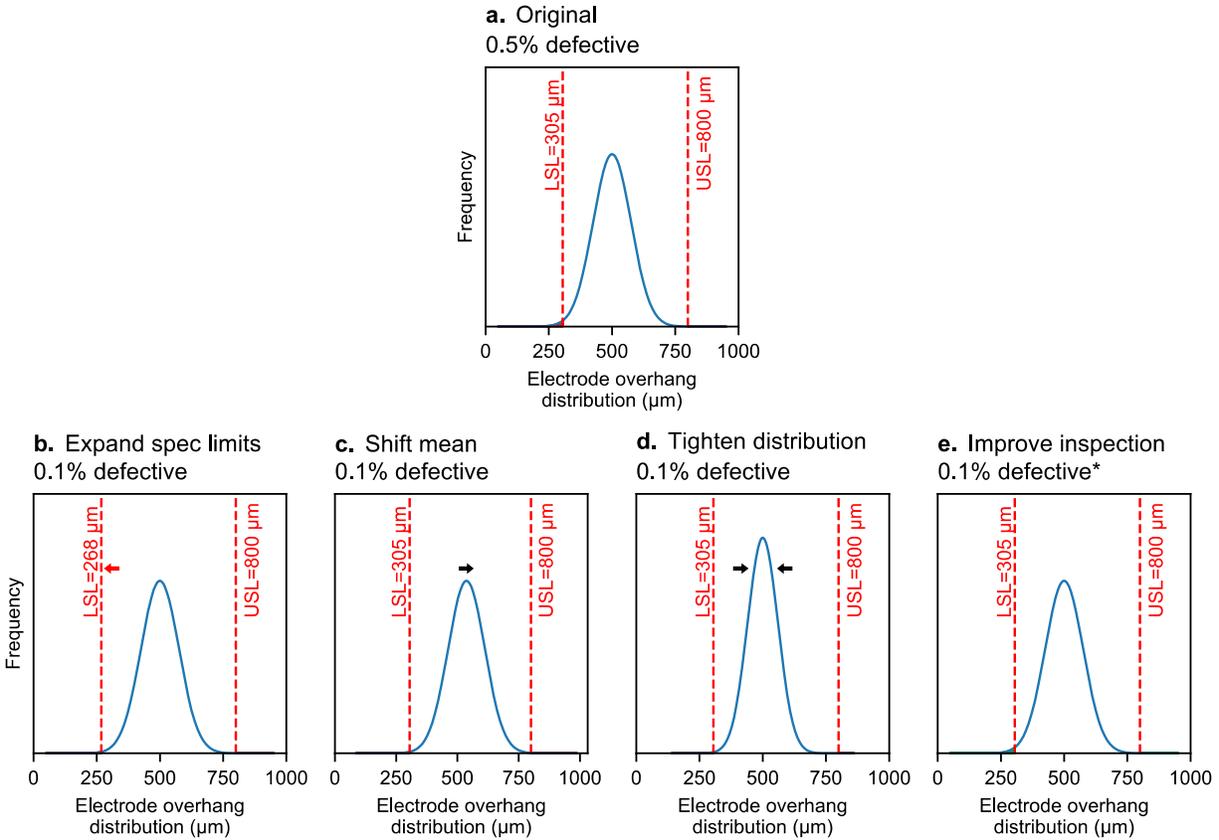

**Figure 5.** Pathways for improving battery quality control, using electrode overhang as an illustrative example. (a) The distribution of electrode overhang for a population of cells, modeled as a Gaussian distribution. The vertical lines represent the lower specification limit (LSL) and upper specification limit (USL), although note that some requirements may only have one of these specification limits. In this example, a fraction (0.5%) of the population falls below the LSL and is thus out of spec. Cell producers can consider four pathways for managing battery quality: (b) expanding the specification limits, (c) shifting the population mean, (d) tightening the distribution, and/or (e) improving inspection. For improved inspection, the 0.1% defect rate assumes an inspection approach in which defective cells do not escape the factory. Note that a defect rate of even 0.1% is much too high (see Figure 3) but exaggerated for illustration purposes. The parameters used to generate these distributions are provided in Supplementary Table I.

Before we begin, we briefly describe key concepts in process capability analysis. A typical requirement has a design target, such as 75 µm for overhang. Each design target has specification limits, or "spec limits", which define the acceptable tolerance range for a given requirement. These spec limits might include a lower spec limit (LSL), upper spec limit (USL), or both. For instance, the LSL and USL for overhang might be 300 µm and 1000 µm (Figure 5a). The motivations for the LSL and USL may differ: in the case of overhang, the LSL would likely be set by reliability or safety concerns (i.e., lithium plating), while the USL might be set by performance or cost concerns (i.e., insufficient positive electrode material and thus low energy, or the cost of excess negative electrode material). The overhang population in production may differ from the production target (e.g., an overhang population with a mean of 500 µm and a standard deviation of 75 µm). A process in violation of its spec limits would be considered a *process control failure*.

***Pathway 1: Expand specification limits (Figure 5b).—***In some ways, the simplest approach for a cell producer to take towards a quality issue is to expand the spec limits. With this approach, no changes are required to the manufacturing process. In some cases, this approach is perfectly valid; for instance, the LSL for overhang could be decreased from, say, 300 µm to 200 µm if the cells were destined for operation in a lower-rate, lower-voltage, and/or lower-lifetime application than their original target use case. However, in many cases, the specifications cannot be changed due to clear safety concerns or even contractual obligations from the cell supplier to its downstream stakeholders.

In other cases, widening specification limits—that is, producing cells that are less reliable and/or safe—can be implemented with coordination from downstream stakeholders. These stakeholders could include module/pack design teams as well as actual end-users. For instance, EV modules and packs are generally designed with some reliability and safety countermeasures in mind.[18,18,19]

A typical pack is passively balanced, which often implies poor resiliency against extreme open-circuit and short-circuit failures.[18,19] In fact, packs with passive balancing are often limited by the weakest cell in a module/pack, where a cell is considered "weak" with regards to its initial energy, remaining energy, or short-circuit current.[15,22,23] Active balancing approaches are employed in some settings, but they add significant cost, weight, and complexity to the pack relative to passive balancing.[16] Additionally, packs are often "over-designed" for safety in that many packs include "extra" thermal insulating materials to prevent a thermal runaway from propagating to adjacent cells, adding additional cost and mass to the pack.[18,19] While this additional safety margin certainly has its benefits, improving cell-level safety could ease requirements for module- and pack-level design, in turn enabling decreased costs and improved performance (i.e., range). Finally, coordinating these changes can be difficult especially if the cell production team and the module/pack design team are from different businesses or organizations and thus have misaligned objectives and/or incentives.

The second type of stakeholder that can be impacted by upstream specification changes is the end user. Specifically, if the cell spec limits are expanded in response to challenges meeting the spec in production, the cell operating limits can be tightened to maintain iso-reliability and -safety to the end user. For instance, if a cell production team is concerned about overhang spec violations, the upper cutoff voltage (which impacts vehicle range) or maximum charge rate of the product could be decreased to maintain similar levels of reliability and safety in operation.[34,48,53] One advantage of this approach is that in extreme cases, the operating limits can be dynamically modified in the field if remote firmware control capabilities are present[91]; however, clear communication with the customer regarding any changes is essential. Of course, the primary downside of this approach is reduced customer utility. In the example above, decreasing the range and/or fast charging time is a highly visible change and will certainly be an unpopular decision among design teams, marketing teams, and customers alike.

Carefully setting specifications is challenging. For instance, the lower specification limit for overhang should certainly be greater than zero, but by how much? The answer depends on not only cell design factors (e.g., electrode thicknesses), production factors (e.g., process capability), and operational factors (e.g., product use case) but also business factors (e.g., financial impacts of yield and throughput metrics, reputational tolerance for quality issues, customer service costs, and more). Furthermore, the timescales of lithium nucleation in the overhang region, lithium nuclei turning into a dendrite, and a dendrite causing a noticeable internal short are worth considering in setting the spec limits but can be difficult to estimate. However, for requirements with potential safety implications, caution is prudent when setting and changing these requirements.

***Pathway 2: Shift the population mean (Figure 5c).***—A second pathway for managing quality is shifting the population mean—in other words, changing the cell design to be more resilient to failure. For example, if a production team were struggling with a wide distribution of overhang, resulting in LSL overhang violations, the design team could agree to increase the overhang design target and the USL, and the production team would respond by increasing the population mean. Unfortunately, this change would lead to lower energy density (since the cell now contains less active material) as well as higher cost per unit energy; designing for high quality and reliability often must come at the expense of performance. Another example of this tradeoff is the use of protective components, such as current interrupt devices, which improve cell safety but add cost and mass.[72] More broadly, new cell chemistries with safety advantages, such as solid-state-batteries, may be even more sensitive to these types of tradeoffs during production due to their increased cell energy and thus increased safety risk.[92] In short, performance and cost are almost always in tension with quality and reliability, and this balancing act is often enormously difficult in light of extreme market pressure to both improve battery performance and reduce cost.

***Pathway 3: Tighten distribution (Figure 5d).—***A third pathway is tightening the distribution of a process parameter, or improving process capability and thus product conformance. In principle, this approach has few downsides: the downstream customers end up with a more uniform product and higher quality without any compromises to performance. In fact, improved conformance can be leveraged for increased performance. In Supplementary Figure 2, we illustrate how increased overhang conformance could translate to a design change that increases cell energy (and thus decreases cost per unit energy as well).

In some cases, cell producers may be able to find low-hanging fruit for improving overall process capability. In practice, however, these types of improvements are often limited by both cost and time. Improving conformance/reducing variability can be an expensive exercise; in the case of overhang, a process engineering team might decide to recalibrate the coating, winding, and/or stacking equipment more frequently, leading to increased operating cost and decreased throughput. Furthermore, for more interdependent failure modes, an understanding of which of the hundreds or even thousands of process parameters may have the biggest impacts on variability is often lacking; even worse, in a process as complex as battery production, changing one parameter will inevitably have downstream effects on another process step or failure mode. Finally, a production team will generally have little interest in tweaking the parameters of an otherwise successful process for an uncertain benefit. In short, reducing variability may be appropriate in some cases but is often impractical in practice. Generally speaking, this approach is counter to design-for-manufacturability principles in which processes should not require narrow process windows to succeed.[93]

***Pathway 4: Improve inspection (Figure 5e).—***The last pathway we examine is improving quality inspection and defect detection during production. As we will discuss, improved inspection may or may not directly change the defect rate depending on the specifics of the approach. While inspection does not fix the root problems on its own, a comprehensive inspection strategy might provide enough relief to prevent some of the most painful trade offs from being made. We recommend McGovern et al.[25] as an excellent review of this topic for a deeper understanding.

We believe that designing an inspection strategy for battery production involves at least three key considerations (Figure 6). Per usual, an "all-of-the-above", case-by-case basis approach is warranted.

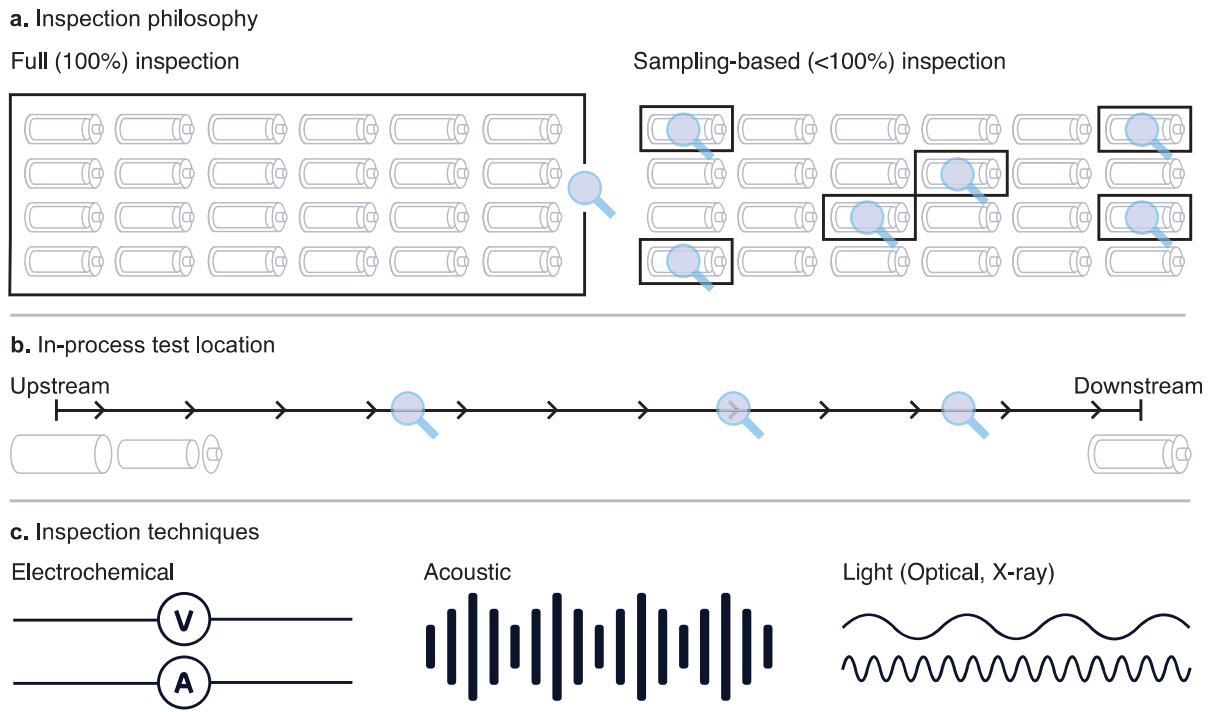

**a.** Inspection philosophy

Full (100%) inspection      Sampling-based (<100%) inspection

**b.** In-process test location

Upstream      Downstream

**c.** Inspection techniques

Electrochemical      Acoustic      Light (Optical, X-ray)

**Figure 6.** Key considerations for designing a strategy for battery quality inspection. (a) Inspection philosophy: either full inspection (100% sampling rate) or sampling-based inspection (<100% sampling rate). (b) In-process test location: either upstream, downstream, or in between. (c) Inspection techniques: either electrochemical, acoustic, or electromagnetic (e.g., visible light or X-rays). Various inspection techniques are further discussed in Table I and Figure 7.

The first consideration is inspection philosophy (Figure 6a). At a high level, two inspection philosophies are full inspection (100% sampling rate) and sampling-based inspection (<100% sampling rate). In full inspection, an inspection test is used as an in-process pass/fail check. Assuming the test is accurate, full inspection obviously prevents defective cells from continuing downstream. Full inspection is often suitable for inexpensive diagnostics where inspecting all or most cells is achievable, such as in-line vision[25], but this approach may add too much operating cost for expensive tests. In contrast, the philosophy of sampling-based inspection is to use the *insights* from inspection tests to root cause issues and estimate the escape rate of defective cells. Sampling-based inspection strategies have been studied and tested for nearly a century and can be quite sophisticated.[94,95] A core assumption of sampling-based inspection is that cell production issues can be traced to one or a couple of suspect process steps or equipment, which is often but not always the case.[96] As a result, careful sampling, monitoring, and analysis can be used to pinpoint many cell failure issues. This approach is often suitable for more expensive diagnostics where 100% detection would add an unacceptably high operating expense. For sampling-based detection, rapid analysis, feedback, and response is essential to ensure that the insights mitigate a small issue from intensifying. In other words, the quality team must remain vigilant to prevent a defective process from remaining defective for days (as opposed to being resolved in hours).

A second consideration for designing an inspection approach is in-process test location (Figure 6b).[97,98] Battery production involves many steps, each of which can introduce new issues. Location optimization must balance the relative advantages of upstream and downstream locations. An upstream test minimizes wasted material and wasted time, as problems can be root caused closer to their source. In contrast, a downstream test maximizes defect detectability because the cell is closer to its final state; for instance, some defects may become readily detectable after formation. Furthermore, cell inspection may continue after the cell has left the production facility. For instance, in addition to the outgoing quality control (OQC) inspection performed by the cell producer, an EV producer may perform incoming quality control (IQC) for incoming cells as the EV producer bears significant reputational risk with a safety incident. We note that in the semiconductor industry, an ensemble of test methods are performed after each process step since every process step can cause an issue. Ultimately, the value of each inspection step must be balanced by its cost.

A final consideration is the inspection tests themselves (Figure 6c). Of course, one of the most important attributes of a test is its ability to detect the defects and features of interest. By extension, the test method must be suitable for its location in the production process. Component tests may be more appropriate for inspecting upstream steps, such as electrode coating; however, testing the nearly-complete product at the end of production may require more advanced characterization techniques for full-cell inspection.

Perhaps the most standard defect detection test performed in battery manufacturing today is measuring the leakage current during rest after formation.[85,99,100] This test can directly capture a key product requirement: no internal shorting. Additionally, the marginal cost of this test is often low since cells are already in electrical fixtures for the formation process. However, this approach has a number of shortfalls intrinsic to any electrochemical test. First, while this test can detect shorts present at the time of the test, it cannot detect latent defects (i.e., defects that will activate sometime after the test and in the field). Second, electrochemical tests measure the global (i.e., non-spatially-resolved) state of the cell, which provides limited diagnostic insight into the root cause of failure. Finally, this test can be very slow (~days). In our opinion, measuring post-formation open-circuit voltage is necessary but insufficient for quality inspection during battery production.

Many other nondestructive techniques can be employed for quality detection in battery production, such as electronic checks (e.g., high-potential testing)[101,102], vision[25], terahertz imaging[25], and acoustic imaging[103,104]. Again, we refer interested readers to McGovern et al.[25] for a deeper understanding. We summarize key attributes of battery quality inspection techniques in Table 1, and in Figure 7 we compare the three nondestructive full-cell imaging techniques — ultrasound (Figure 7a), 2D X-ray (Figure 7b), and 3D X-ray (Figure 7c) (the experimental details are discussed in Supplementary Discussion 2). In our view (although please see our conflict of interest statement in the Acknowledgements section), X-ray inspection, specifically 3D X-ray inspection (computed tomography), provides exceptionally rich insights into battery quality[62,105,106] with a clear pathway for high scalability[107–110]. Ultimately, however, we believe an arsenal of characterization techniques is the best defense against battery quality issues in production.

**Table I.** Key requirements of cell-level battery quality inspection techniques. Materials-level characterization techniques, such as electron microscopy, are excluded from this table. A test duration of ≤10 s is important so that a meaningful number of cells (≥~10k/day) can be sampled from production. Spatial resolution on the order of 10–100 μm is important for detecting many critical battery defects, such as electrode overhang[53,74,75] and metallic particle contaminants[26,27,76]. In our opinion, an ensemble of complementary inspection techniques is the best defense against battery quality issues.

| Technique | Non-destructive | Scalable to ≤10 s/cell | Full cell inspection | Spatially resolved | Resolution of ≤50 μm |
|---|---|---|---|---|---|
| Cycling and storage | No | No | Yes | No | N/A |
| Ultra High Precision Coulometry (UHPC) | No | No | Yes | No | N/A |
| Formation checks (e.g., OCV decay) | Yes | No | Yes | No | N/A |
| Electrochemical Impedance Spectroscopy (EIS) | Yes | No | Yes | No | N/A |
| High Potential testing (HiPot) | Yes | Yes | Yes | No | N/A |
| Dissection | No | No | Yes | Yes | Yes |
| Cross section | No | No | No | Yes | Yes |
| In-line vision | Yes | Yes | No | Yes | Yes |
| Acoustic imaging | Yes | Yes | Yes | Yes | No |
| 2D X-ray imaging | Yes | Yes | No | Yes | Yes |
| 3D X-ray imaging (CT) | Yes | Yes | Yes | Yes | Yes |

**a. Ultrasound**

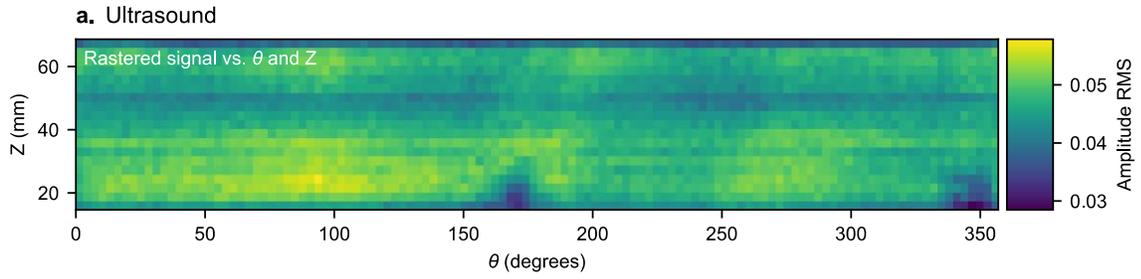

**b. 2D X-ray**

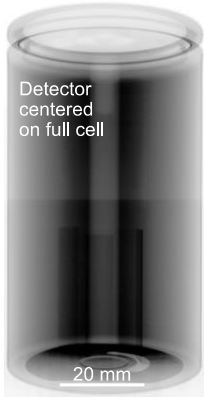

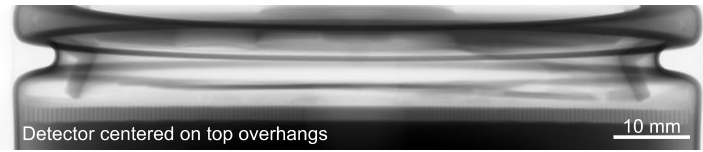

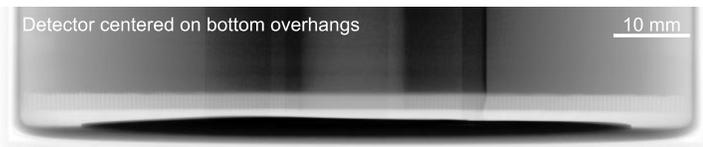

**c. 3D X-ray (CT)**

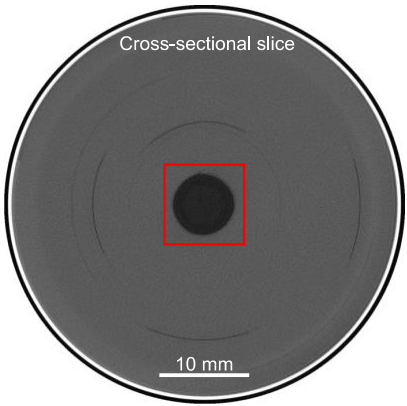

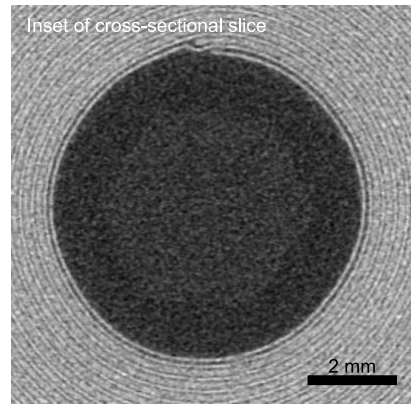

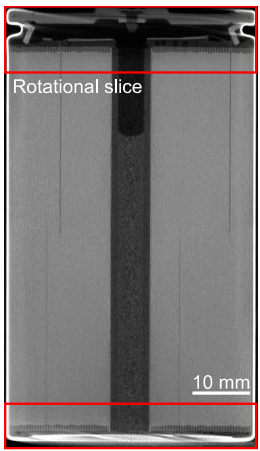

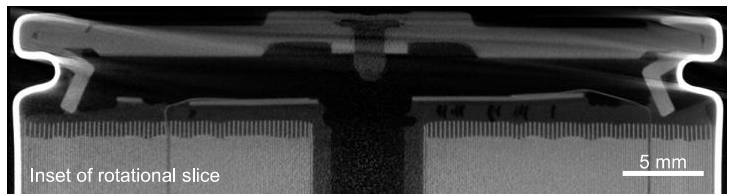

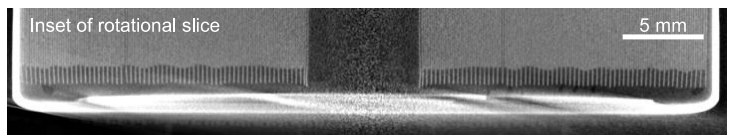

**Figure 7**. Comparison of nondestructive, full-cell, spatially-resolved techniques for evaluating battery quality. All measurements were obtained on a BYD FC4680 cylindrical cell. Experimental details are provided in Supplementary Discussion 2. (a) Ultrasound (acoustic) imaging. The root-mean-square (RMS) value of the amplitude is displayed as a function of $Z$ (position along the cell height) and $\theta$. The acquisition time was three hours, although this time could be reduced with a multi-transducer pair system as opposed to the single transducer pair system used here. (b) 2D X-ray imaging. Three distinct images are presented: an image of the full cell, an image centered on the top overhang region, and an image centered on the bottom overhang region. The acquisition time for each image was 125 milliseconds. (c) 3D X-ray imaging (computed tomography). Two different slice orientations, along with insets of key regions, are displayed. The acquisition time was 73 seconds.

Inspection tests during production can generate massive quantities of data.[111,112] These data can serve as a continuously updated snapshot into battery quality if carefully organized and managed—and especially if combined with data from the manufacturing process. As previously discussed, dynamicism is a key ingredient to manufacturing success; the data from inspection tests enable process resiliency in that engineering, production, and quality teams can quickly understand the impact of a process change on cell quality. In particular, these records are invaluable in case of a field failure event to evaluate the size of the affected cell population years after the cells were produced. Data analytics and artificial intelligence tools for anomaly detection and correlation analysis could be powerful aids to glean insights from this data.[111,112]

Lastly, we discuss the dependence of quality and defect detectability on form factor. The battery industry is currently pursuing three primary form factors: cylindrical, pouch, and prismatic. While many design criteria influence the "optimum" form factor for a given application, we propose that both quality and "quality inspectability" are also important. Currently, the industry lacks a clear view of the relationship between form factor, quality, and quality inspectability. Major considerations for quality include wound vs. stacked, the strength of the casing (with pouch having the least rigid case), and intrinsic heterogeneities from intra-cell thermal, mechanical, and (electro)chemical gradients.[113,60,114–116] Major considerations for detectability are overall size and aspect ratios (i.e., is a cell geometry "2D", such as a pouch cell, or "3D", such as a cylindrical cell). Lastly, for EVs, the number (and arrangement) of cells in a pack is an important factor for pack reliability (see Figure 2).[14,17] Form factors that generally house higher-energy cells, like prismatic, typically have hundreds of cells per pack; form factors with smaller cells, like cylindrical, typically have thousands of cells per pack.[24] In this case, the optimum form factor from a quality standpoint depends on if the most concerning defects occur primarily per unit of energy or per cell. Different defects will have different dependencies. For instance, electrode-level defects will likely occur per unit of energy, and thus more cells will fail for higher-energy form factors; conversely, cell-level defects like weld issues will likely see higher failure rates for lower-energy form factors.

## Conclusions

The need for planetary-scale battery production has never been clearer, and cell suppliers around the globe are racing to ramp capacity. As we hope we have convincingly argued, however, ensuring high quality during battery production is an immensely challenging endeavor with enormous financial, reputational, environmental, and human stakes. Yet the battery quality challenge is underappreciated given its criticality to global electrification efforts. We believe this problem is of similar importance to improving battery performance, the predominant focus of academic and industrial battery engineering today; in fact, in many ways, performance and quality are closely intertwined and often come into conflict. Both cell producers and OEMs must properly weigh both factors in cell design and selection.

The industry has a long road ahead to enable successful management of battery quality. We believe the following steps would be immensely beneficial for the industry. First, a complete standardized taxonomy of battery failure modes would enable a common language and understanding around battery quality issues. Similarly, a library of tests for each failure mode would enable rapid qualification and issue resolution throughout the industry. While some standard tests exist today[117], a much larger test library is needed to match the wide variety of failure modes and requirements. Additionally, modeling tools to answer critical quality questions (e.g., where to set the spec limits for overhang) will be highly beneficial for the industry. Furthermore, faster, less expensive, and more information-rich battery quality characterization techniques are sorely needed to quickly test the massive quantities of cells produced daily at a typical cell production facility—along with user-centric analytics tools to turn this massive volume of data into actionable insights. Lastly, stronger efforts around preventing counterfeit and low-quality cells, as well as developing stronger quality and safety certifications, will strengthen the reputation of battery-employing products. Overall, we believe that a collaborative industry-wide effort to improve battery quality would bolster investor, legislative, regulatory, and customer confidence in this technology; conversely, every publicly reported safety event threatens the success of the entire industry.

Battery quality also has important impacts for questions around battery reuse and recycling.[118] While energy retention is an important metric to determine suitability for reuse, the presence of cell failure and defects is arguably the primary gating item for this decision. Cost-effective characterization techniques for battery quality may enable more cells to be reused in a second-life application instead of immediately recycled.[78]

While regulatory efforts around battery safety are only just beginning[13], we expect more will arise as more battery safety incidents occur. Existing proposals such as the EU's "battery passport" include a quality component.[119,120] In our opinion, new regulations should not impede creativity and progress in this critical and burgeoning industry, but they should provide sensible safeguards to protect public safety. In particular, preventing low-cost, poor-quality cells from entering the market would help avert the worst of these safety issues from occurring.[10,71]

We hope our perspective provides a small step towards a safely-electrified future.

## Acknowledgements

We thank Amariah Condon, Michael Chen, and Andrew Weng for insightful discussions, Kristen Giuliano for designing many of the schematics, and Shaurjo Biswas (Liminal) for the ultrasound data. The authors are all co-founders with a financial interest in Glimpse Engineering Inc., which specializes in high-throughput CT scanning for batteries.

## Code availability

All data and code used to generate the figures is available at https://github.com/petermattia/battery-quality-at-scale.

## CRediT author contributions statement

**Peter Attia:** Conceptualization, Writing - Original Draft, Writing - Review & Editing, Visualization.
**Eric Moch:** Writing - Review & Editing. **Patrick Herring:** Writing - Review & Editing.